\documentclass{ws-p8-50x6-00}

\begin{document}

\title{QCD-improved factorization in nonleptonic $B$ decays}

\author{Junegone Chay}

\address{Department of Physics, Korea University, Seoul 136-701, Korea
\\E-mail: chay@korea.ac.kr}

\maketitle

\abstracts{I consider nonleptonic decays of $B$ mesons into two light
mesons using the light-cone wave functions for the mesons. In the
heavy quark limit, nonfactorizable contributions are calculable from
first principles in some decay modes. I review the idea of 
the QCD-improved factorization method and discuss the implications in
phenomenology.} 

\section{Introduction}
Nonleptonic decays of $B$ mesons have attracted a lot of attention
recently since they were observed experimentally in CLEO, BaBar and
BELLE. These decays are important in extracting the information on the 
Cabibbo-Kobayashi-Maskawa (CKM) matrix elements and CP
violation. On the theoretical side, it is the least understood
area. Since nonleptonic decays involve nonperturbative effects 
such as final-state interactions, it is difficult to obtain a
quantitative theoretical prediction. However, it has been recently
found that nonperturbative effects such as the nonfactorizable
contribution in nonleptonic decays could be systematically understood
in the heavy quark limit with $m_b \rightarrow \infty$. 

Here I will review the current
status of understanding on nonleptonic $B$ decays very
schematically. I will focus on the underlying ideas omitting technical
complication. I hope that this talk will give a clear sketch of what
is being studied currently. 
It has been found that nonfactorizable contributions in
nonleptonic $B$ decays into two light mesons could be calculated using 
perturbation theory in the heavy quark limit $m_b \rightarrow
\infty$. I will explain in detail how we treat nonleptonic $B$ decays
in the heavy quark limit and discuss some phenomenological aspects. 

The theoretical framework for $B$ decays is to use the effective weak
Hamiltonian at the renormalization scale $\mu \approx m_b$, which is
schematically given as
\begin{equation}
H_{\mathrm{eff}} = \frac{G_F}{\sqrt{2}} \sum_q V_{qb} V_{qd}^* C_i
(\mu) Q_i^q (\mu), 
\end{equation}
where $V_{ij}$ are the CKM matrix elements, $C_i$ are the Wilson
coefficients and $O_i^q$ are the effective four-quark operators. The
Wilson coefficients are calculable order by order in perturbation
theory, and the main issue in analyzing nonleptonic decays is how to
evaluate the matrix elements of the four-quark operators. For example,
if $\overline{B}$ decays into the final two mesons $M_1$ and $M_2$,
how do we evaluate $\langle M_1 M_2 | O_i^q |\overline{B}\rangle$?

First we can use the naive factorization in which we neglect the
strong interaction effects between mesons and separate the operators
into a current-current form, and calculate their matrix
elements \cite{bsw}. Schematically this process can be expressed as
\begin{equation}
\langle M_1 M_2 | O_i |B\rangle \approx \langle M_1 |J_{\mu} |0\rangle
\langle M_2 |J^{\mu}|B\rangle.
\end{equation}
Here $J_{\mu}$ is the current operator in the effective
Hamiltonian. Each matrix element is parameterized by a decay constant
or a form factor, which describe intrinsic nonperturbative
effects. The decay constant and the form factors can be obtained from
either experiment or other theoretical techniques such as QCD sum
rules \cite{sumrule}. 

Unfortunately this naive factorization is unsatisfactory. First of
all, there is no justification in neglecting the final-state
interactions between mesons. The argument of color transparency
\cite{bj} can be applied in the case of two final light mesons, but
it should be proved explicitly if the matrix elements can be truly
factorized including final-state interactions. Secondly, theoretically
the naive factorization gives an unphysical result. Decay constants
and form factors are independent of the renormalization scale $\mu$,
and the decay amplitude has an arbitrary $\mu$ dependence through the
Wilson coefficients $C_i (\mu)$. This is because we replace the matrix
elements, which depend on $\mu$, by the decay constants and the form
factors, which are independent of $\mu$. Therefore the resulting
amplitude is unphysical. 

As a remedy to this problem, Ali and Greub \cite{ali} suggested to
calculate the radiative corrections of the operators before taking the
matrix elements. Following this procedure, we can write the matrix
element as
\begin{equation}
\langle M_1 M_2 | O_i |B\rangle \approx F(\mu, \alpha_s) \Bigl[
\langle M_1 |J_{\mu}|0\rangle \langle M_2 |J^{\mu}
|B\rangle\Bigr]_{\mbox{tree}}. 
\end{equation}
It turns out that the $\mu$ dependence in $F(\mu, \alpha_s)$ arising
from the radiative corrections cancel the $\mu$ dependence in the
corresponding Wilson coefficients at any given order, hence the decay
amplitude does not depend on the renormalization scale. I call this
procedure the ``{\em improved factorization}''. 

The unphysical $\mu$ dependence is absent in the improved
factorization, but it has other significant problems. First of all, in
order to avoid infrared divergence, $F(\mu, \alpha_s)$ is calculated
with off-shell external quarks \cite{buras}. The external momenta
$p^2$ play a role of the infrared cutoff. However, because of this,
the decay amplitudes depend on the choice of the gauge, and the
employed regularization schemes \cite{buras2}. Therefore the decay
amplitude depends on arbitrary gauges and the schemes, hence also
unphysical. 

If we put external quarks on their mass shell, the gauge dependence
and the scheme dependence go away, but, in this case, the amplitude
becomes infrared divergent and gives also an unphysical
result. Therefore we need a consistent scheme which solves all the 
problems I mentioned above. Before I explain the main idea of the
QCD-improved factorization, note that the Feynman diagrams from which
infrared divergence appears are those with vertex corrections of the
weak currents. 

\section{QCD-improved factorization}

The source of infrared divergence in the radiative corrections of the
effective weak Hamiltonian, as mentioned above, suggests an
interesting idea. The Feynman diagrams which cause infrared divergence
are the radiative corrections of vertices. In other words, the infrared
divergence comes from the radiative
corrections for the decay constant of a light meson and form factors
for $B\rightarrow M_2$. 

It reminds us of the hadron-hadron scattering, in which the infrared
divergence of the scattering amplitude is absorbed in the redefinition
of the parton distribution functions. The remaining hard scattering
amplitude and the parton distribution functions are factorized. We can
apply the same idea to $B$ decays. The infrared divergences can be
attributed to the renormalization of the decay constant and the form
factors, and other radiative corrections constitute nonfactorizable
contributions. That is, the infrared divergence is absorbed in the
definition of the light-cone meson wave function or the form
factors. This is first observed in Ref.~\cite{politzer}

Recently Beneke et al. \cite{beneke} formulated this problem in the
heavy quark limit and extensively studied nonleptonic $B$ decays into
two final-state mesons. The idea can be summarized as follows:
We first take the heavy quark limit $m_b \rightarrow \infty$. In this
limit we can calculate nonfactorizable contributions systematically in
perturbative QCD. Also we can obtain corrections to the perturbative
results by expanding in powers of $\Lambda_{\mathrm{QCD}}/m_b$. 

The next step is to
arrange external quarks using Fierz transformation such that the
quark-antiquark pair, which forms a meson, is included in a single
current. And we use the light-cone wave function for the mesons. If we
can use the operators in the effective Hamiltonian as 
they are in arranging quarks, we call that configuration of quarks as
``charge-retention configuration''. If we have to switch some quarks
using Fierz transformation, we call that configuration as
``charge-changing configuration''. This arrangement is important since
it determines which radiative corrections correspond to the
corrections to the decay constants or the form factor. Since the decay
amplitude in each process depends on how we arrange quarks, the
scattering amplitude becomes process-dependent. 

The infrared divergence is attributed to the renormalization of the
wave function or the form factors. And we calculate all the 
nonfactorizable contributions along with the spectator contribution
and the annihilation channels. This procedure is called the
``{\em QCD-improved factorization}''. If all the nonfactorizable
contributions are infrared finite, and suppressed as $m_b$ goes to
infinity, these processes can be treated in the QCD-improved
factorization. In this case, we have a theoretical method to analyze
nonleptonic $B$ decays from first principles. 
In the QCD-improved factorization method, we can formally write the
matrix element as 
\begin{equation}
\langle M_1 M_2 | O_i |B\rangle \approx \langle M_1 |J_{\mu}|0\rangle
\langle M_2 |J^{\mu}|B\rangle \Bigl[ 1 + O(\alpha_s)
+O(\Lambda_{\mathrm{QCD}}/m_b) \Bigr].
\end{equation}
If we neglect the radiative corrections and the $1/m_b$ corrections,
we restore the result obtained in the naive factorization. The matrix
element can be written as a product of a decay constant and a form
factor. And we can systematically calculate the corrections to the
result in the naive factorization.

Before we discuss the details of the QCD-improved factorization, I
would like to explain another approach to study nonleptonic $B$
decays. Keum et al. \cite{keum} have also considered nonleptonic $B$
decays using light-cone wave functions, but they
concentrated on the calculation of the form factors instead of
nonfactorizable contributions. They calculated a single gluon exchange
which is responsible for the correction to the form factor. There also
appears infrared 
divergence in the form factor, but they introduce the
Sudakov factor which smears the endpoint region such that there is no
infrared divergence. And the origin of imaginary parts in their
calculation comes from the modification of the propagator with
transverse momentum, which is totally different from the source of the
imaginary part in the QCD-improved factorization.

They do not consider nonfactorizable contributions at the moment, and
hence the leading-order Wilson coefficients are employed. Because what
is calculated is totally different in the two approaches, we have to
be careful when we try to compare the results in both approaches. 

\section{Nonfactorizable contribution}
Let us go into the detail of how to calculate nonfactorizable
contributions. When we use the light-cone meson wave functions for
exclusive decays, the transition amplitude of an operator $O_i$ in the
weak effective Hamiltonian is given by
\begin{eqnarray}
\langle M_1 M_2 | O_i|\overline{B}\rangle &=& \sum_j
F_j^{B\rightarrow M_2}  \int_0^1 dx T_{ij}^I (x) \phi_{M_1}
(x) \nonumber \\
&&+ \int_0^1 d\xi dx du T_i^{II} (\xi, x,u) \phi_B (\xi) \phi_{M_1}
(x) \phi_{M_2} (u),
\label{imqcd}
\end{eqnarray}
where $F_j^{B\rightarrow M_2}$ are the form factors for
$\overline{B}\rightarrow M_2$, and $\phi_{M_i} (x)$ is the
light-cone wave function for the meson $M_i$. $T_{ij}^I (x)$ and
$T_i^{II} (\xi, x,u)$ are hard-scattering amplitudes, which are
perturbatively calculable. The second term in Eq.~(\ref{imqcd})
represents spectator contributions.

\begin{figure}[h]
\vspace{0.5cm}
\begin{picture}(200,70)(10,5)
\unitlength=0.65mm
\thicklines
\put(0,20){\line(1,0){40}}
\put(0,20){\vector(1,0){10}}
\put(20,20){\vector(1,0){10}}
\put(50,20){\line(1,0){40}}
\put(50,20){\vector(1,0){10}}
\put(70,20){\vector(1,0){10}}
\put(100,20){\line(1,0){40}}
\put(100,20){\vector(1,0){10}}
\put(120,20){\vector(1,0){10}}
\put(150,20){\line(1,0){40}}
\put(150,20){\vector(1,0){10}}
\put(170,20){\vector(1,0){10}}
\put(20,20){\circle*{2}}
\put(70,20){\circle*{2}}
\put(120,20){\circle*{2}}
\put(170,20){\circle*{2}}

\put(5,40){\line(1,-1){15}}
\put(5,40){\vector(1,-1){7.5}}
\put(20,25){\line(1,1){15}}
\put(20,25){\vector(1,1){7.5}}
\put(20,25){\circle*{2}}

\put(55,40){\line(1,-1){15}}
\put(55,40){\vector(1,-1){7.5}}
\put(70,25){\line(1,1){15}}
\put(70,25){\vector(1,1){9}}
\put(70,25){\circle*{2}}

\put(105,40){\line(1,-1){15}}
\put(105,40){\vector(1,-1){6}}
\put(120,25){\line(1,1){15}}
\put(120,25){\vector(1,1){7.5}}
\put(120,25){\circle*{2}}

\put(155,40){\line(1,-1){15}}
\put(155,40){\vector(1,-1){7.5}}
\put(170,25){\line(1,1){15}}
\put(170,25){\vector(1,1){12}}
\put(170,25){\circle*{2}}

\multiput(14.5,29)(-3,-3){4}{\oval(3,3)[tl]}
\multiput(11.5,29)(-3,-3){3}{\oval(3,3)[br]}

\multiput(75.5,29)(3,-3){4}{\oval(3,3)[tr]}
\multiput(78.5,29)(3,-3){3}{\oval(3,3)[bl]}
\multiput(111.5,32)(5,-3){5}{\oval(5,3)[tr]}
\multiput(116.5,32)(5,-3){4}{\oval(5,3)[bl]}
\multiput(178.5,32)(-5,-3){5}{\oval(5,3)[tl]}
\multiput(173.5,32)(-5,-3){4}{\oval(5,3)[br]}

\put(2,12){$b$}

\put(52,12){$b$}

\put(102,12){$b$}

\put(152,12){$b$}
\end{picture}
\vspace{-0.5cm}
\caption{Feynman diagrams for nonfactorizable contribution at order
$\alpha_s$. The dots represents the operators $O_i$.}
\label{fig1}
\end{figure}
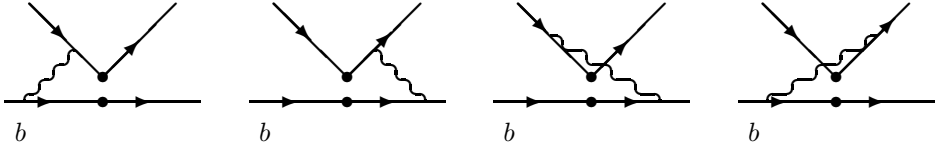

The relevant Feynman diagrams for $T_{ij}^I$ are shown in Fig.~1. 
Each Feynman diagram has an infrared divergence. But if we sum over
all the Feynman diagrams and symmetrize with respect to
$x\leftrightarrow 1-x$, where $x$ is the momentum fraction of the
outgoing meson, we have an infrared-finite result. Another feature of
the nonfactorizable contribution is that there appears an imaginary
part due to the final-state interactions. This plays an important role
in studying CP violation in nonleptonic decays. The strong phase is
calculable in perturbation theory. And since we are working at
next-to-leading order accuracy, the dependence on $\mu$ becomes very
mild. 

\begin{figure}[h]
\vspace{0.5cm}
\begin{picture}(100,70)(-80,0)
\unitlength=0.6mm
\thicklines
\put(0,20){\line(1,0){40}}
\put(50,20){\line(1,0){40}}
\put(0,10.5){\line(1,0){40}}
\put(50,10.5){\line(1,0){40}}
\put(20,20){\circle*{2}}
\put(70,20){\circle*{2}}

\put(5,40){\line(1,-1){15}}
\put(20,25){\line(1,1){15}}
\put(20,25){\circle*{2}}

\put(55,40){\line(1,-1){15}}
\put(70,25){\line(1,1){15}}
\put(70,25){\circle*{2}}
\multiput(13.5,30)(0,-6){4}{\oval(3,3)[l]}
\multiput(13.5,27)(0,-6){3}{\oval(3,3)[r]}
\multiput(76.5,30)(0,-6){4}{\oval(3,3)[r]}
\multiput(76.5,27)(0,-6){3}{\oval(3,3)[l]}
\end{picture}
\caption{Feynman diagrams for the spectator contribution.}
\label{fig2}
\end{figure}
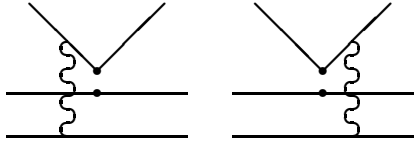

There are other nonfactorizable contributions such as the spectator
contributions. The Feynman diagrams for $T_{ij}^{II}$ are shown in
Fig.~2. In calculating nonfactorizable contributions, we use the
light-cone wave functions. The projection of the quark bilinears to
each pseudoscalar light meson wave function to the order of twist
three can be written as 
\begin{eqnarray}
&&\langle P (p) | \overline{q}_{\alpha} (y) q^{\prime}_{\beta} (x)
|0\rangle = \frac{if_P}{4}  \int_0^1 du e^{i(u p\cdot y + (1-u)
p\cdot x)} \nonumber \\
&&\times \Bigl\{ \not{p} \gamma_5 \phi (u) -\mu_P \gamma_5 \Bigl(
\phi_p (u) -\sigma_{\mu\nu} p^{\mu} (y-x)^{\nu}
\frac{\phi_{\sigma}(u)}{6} \Bigr) \Bigr\}.
\end{eqnarray}
For the $B$ meson, we can write the projection as
\begin{equation}
\langle 0| \overline{q}_{\alpha} b_{\beta} |\overline{B}\rangle =
-\frac{if_B}{4} \phi_B (\xi) \Bigl\{ (\not{p}_B + m_B) \gamma_5
\Bigr\}_{\beta\alpha}.
\end{equation}
Here $\phi$ is the leading-twist wave function, and $\phi_p$ and
$\phi_{\sigma}$ are twist-three wave functions for the pseudoscalar
and the tensor currents respectively. For the $B$ wave function, we
take the leading-twist wave function only. Since we expand in powers
of $1/m_b$, at leading order $\phi_B (\xi) \sim \delta (\xi
-\Lambda_{\mathrm{QCD}}/m_b)$. In calculating the spectator
contribution, we have the integral of the form
\begin{equation}
\int_0^1 d\xi \frac{\phi_B (\xi)}{\xi} \approx
\frac{m_B}{\Lambda_{\mathrm{QCD}}}, 
\end{equation}
which is enhanced.

We can consider corrections of order
$O(\Lambda_{\mathrm{QCD}}/m_b)$. The most important contribution comes
from the term proportional to $\mu_P/m_b$ which is given by
\begin{equation}
\mu_P = \frac{m_P^2}{m_1 + m_2},
\end{equation}
where $m_1$ and $m_2$ are the masses of the quarks which form a meson
with mass $m_P$. Compared to other $O(\Lambda_{\mathrm{QCD}})$ terms,
this is numerically large. For the case of $\pi^+$, for instance, it
is $\mu_{\pi^+} \sim 1.4 $ GeV. Therefore it has been of great interest
to calculate higher-twist effects proportional to $\mu_P$. However,
the spectator contribution from $O_1$ and $O_2$ with higher-twist wave
functions is infrared divergent \cite{beneke2}. 
It has been suggested that we introduce some parameters to regulate
the infrared divergence and regard the parameter as a theoretical
uncertainty. But these contributions are numerically
large and especially the extraction of the strong phase becomes too
ambiguous to say anything quantitatively.

The annihilation topology poses another problem. When we calculate the
spectator contribution with the operator $O_5$, the amplitude is also
infrared divergent. Therefore the annihilation topology can give a
significant power correction to the decay amplitude. The analysis on
the annihilation channel with radiative correction is in progress.

One way to look at the infrared divergence is that it is not a serious
problem. The divergence comes from the endpoint and it actually gives
the logarithmic enhancement as
\begin{equation}
\int_{\Lambda_{\mathrm{QCD}}/m_b}^1 \frac{du}{u} \approx \ln
\frac{m_b}{\Lambda_{\mathrm{QCD}}}.
\end{equation}
Since the cutoff $\Lambda_{\mathrm{QCD}}$ is actually arbitrary, we
can parameterize this contribution and regard this as a theoretical
uncertainty. But its magnitude is numerically large, and thus it
enlarges theoretical uncertainty. 

Another view is a more conservative one. If there appears an infrared
divergence in the hard scattering amplitude, the effect of soft gluon
exchange is really significant and the QCD-improved factorization
breaks down at this order. It remains to be seen if the QCD-improved
factorization really breaks down, or there are some other
contributions which cancel the infrared divergence rendering the final
result infrared finite.

\section{Conclusion}

The understanding of nonleptonic $B$ decays into two mesons has
acquired a new sophisticated level. In the heavy quark limit,
nonfactorizable contributions are calculable using perturbative QCD
for light final-state mesons. When one of the final-state meson is
heavy, we can still use the QCD-improved factorization for the case in
which the spectator quark in the $B$ meson goes to the heavy meson in
the final state. If the spectator quark goes to a light meson, as in
class II decays, the nonfactorizable contribution is infrared
divergent, and the effect of soft gluon exchange is significant. 

But the analysis of higher-twist effects is yet far from
satisfactory. For example, the spectator contribution which is
proportional to the twist-three contribution of the meson wave
function is infrared divergent. And the annihilation topology also has
the infrared divergence. The status of the QCD-improved factorization
method for nonleptonic $B$ decays into two mesons is not complete
until we disentangle the infrared divergence to give a quantitative
prediction. 

It will be an interesting project to combine the QCD-improved
factorization with the calculation of the form factors using the
light-cone wave functions. Currently, in the QCD-improved
factorization, we use the form factors extracted from experiment, as
in semileptonic $B$ decays. On the other hand, in Ref.~\cite{keum}
they only consider the calculation of form factors. It will be
interesting to see whether we can give a consistent theoretical
description of nonleptonic $B$ decays combining these two approaches. 

\section*{Acknowledgments}
I would like to thank Pyungwon Ko for the collaboration on the
subject. I am also grateful to the organizers of the conference for
their kind support. This research is supported in part by the Ministry
of Education grants KRF-99-042-D00034 D2002, the Center for Higher
Energy Physics, Kyungbook National University, and Korea University.

\end{document}